# Electrochemical Synthesis of Superconductive Boride $MgB_2$ from Molten Salts

Hideki Abe*[1] and Kenji Yoshii*[2]

[1]*National Institute for Material Science (NIMS), Tsukuba, Ibaraki 305-0047, Japan*
[2] *Japan Atomic Energy Research Institute (JAERI), Mikazuki, Hyogo 679-5143, Japan*

Electrolysis has been performed on a fused mixture of magnesium chloride, potassium chloride, and magnesium borate at 600 $^o$C under an Ar atmosphere using a graphite anode and a Pt cathode. Magnetic measurements and X-ray diffractometry have revealed that the resultant deposit on the cathode contains a few tens molar percent of $MgB_2$. A voltanmetric measurement at the initial stage of electrolysis has shown that the threshold cathode voltage for the formation of $MgB_2$ from the molten salts is -1.6 V.

Since the discovery of superconductivity in $MgB_2$ at 39 K, [1] we have learned that this compound possesses a potential for superconductive materials for high-performance electronic devices or high-power magnets. Its critical temperature for superconductivity ($T_c$) is almost twice as high as $T_C$ = 23 K of $Nb_3Ge$, which has been the highest critical temperature among the known intermetallic superconductors. [2] A remarkably high critical current density ($J_c$) with an order of $10^5$ $Acm^{-2}$ under a field of 10 T at 4.2 K has been achieved by oxygen alloying. [3] The low cost of the constituent elements, Mg and B, is also an advantage of $MgB_2$ compared to the Nb-based superconductive materials. For the reason above, it is expected that practical superconductor products, in particular superconducting wires with reasonable prices and high operating temperatures, will be made out of $MgB_2$.

The first report on the identification of $MgB_2$ goes back a half century, at which time $MgB_2$ was prepared by a direct reaction between powdered B and excessive Mg in a MgO-coated Fe crucible at 800 $^o$C under a hydrogen flow.[4] Recent refined preparation methods of wires and single crystals are also based on the direct reaction between the elements.[5-7] $MgB_2$ wires have been fabricated by exposing thin wires of B to high-pressure Mg vapor at 950 $^o$C in a sealed Ta ampoule.[5,6] Single crystalline $MgB_2$ has been prepared in a similar way at high temperatures from 1000 $^o$C to 1400 $^o$C in a sealed Mo crucible containing Mg and B chunks.[7] In both cases, $MgB_2$ is synthesized through the reaction between high-pressure Mg gas and solid B in enclosed systems at temperatures near the boiling point of Mg, 1090 $^o$C. Because of the high volatility of Mg and the high melting point of B, such severe conditions for preparation are unavoidable as long as the elements Mg and B are chosen as the starting materials.[8]

On the other hand, it has been known that several borides containing volatile elements such as alkaline-earth or rare-earth elements can be synthesized by electrolysis on fused mixtures of halides and borates.[9-11] It is astonishing that single crystals of $LaB_6$ with an extremely high melting temperature of 2500 $^o$C can be prepared by electrolysis on molten salts at temperatures lower than 1000 $^o$C.[9,10] In particular, the electrochemical preparation of $CaB_6$ from a molten mixture of $CaCl_2$ and $CaB_4O_7$ has suggested the possibility of the synthesis of $MgB_2$ from fused $MgCl_2$ and $MgB_2O_4$ in a similar manner.[11]

In this paper, we report the results on the electrolysis of a fused mixture of $MgCl_2$, KCl, and $MgB_2O_4$. Magnetic measurements and X-ray diffractometry indicate that the resultant deposits on the cathode contain a few tens molar percent of $MgB_2$.

Figure 1(a) shows a top view of the electrolysis cell before the electrolysis procedure. An alumina boat with sizes of 10 x 10 x 100 mm$^3$ in width, depth, and length, respectively, was used for the electrolyte container. A graphite rod with a diameter of 6.0 mm was placed at one end of the boat as an anode (denoted by A), and a Pt plate with 0.1 mm thickness and 5 mm width was fixed at the other end as a cathode (denoted by B). A Pt wire with a diameter of 1.0 mm was set at the middle of the boat as a reference pole for voltanmetric measurements (denoted by C). All of the electric poles (anode, cathode and reference pole) were connected to Pt leads with a diameter of 0.1 mm. The electrolysis cell was filled with 2.0 g electrolyte, which was composed of $MgB_2O_4$, KCl, and $MgCl_2$ with a molar ratio of 1 : 5 : 5. KCl was added to the electrolyte in order to lower the melting point and the viscosity of the fusion. The electrolysis cell was placed in a tubular furnace with an insertion quartz tube with a diameter of 40 mm. One end of the insertion tube was joined to a dry Ar gas line, and the other end was connected to an exhaust. The Pt leads were led out of the insertion tube via a hermetic connector to a DC electric power supply outside.

First of all, the temperature of the furnace was raised from room temperature to 400 $^o$C and kept for 2 hours under an Ar gas flow, in order to dry the electrolyte. The temperature was then raised to 600 $^o$C, at which the electrolyte powder turned into a colorless transparent fusion. After waiting 30 minutes for the stabilization of the fusion, a constant DC voltage of 5 V was applied between the anode and cathode, and the electrolysis system was kept for 2 hours under the conditions. During the electrolysis procedure, the amplitude of the electric current flowing from the anode to the cathode decreased gradually from the initial amplitude of several tens milli-Ampere to an order of milli-Ampere. After the electrolysis procedure, the temperature of the furnace was lowered down to room temperature, and the electrolysis cell was taken out of the tubular furnace.

Figure1(b) shows a top view of the electrolysis cell after the electrolysis procedure. The bottom of the cell is covered with the crystallized electrolyte. A black slug-shaped deposit (denoted by D) grows from the tip of the cathode (B), whereas no precipitate is recognized on the graphite anode (A) or on the Pt reference pole (C). The black deposit (D) was removed from both the bottom of the electrolysis cell and the Pt cathode mechanically and was characterized by magnetic measurements using a SQUID magnetometer (SPMS, Quantum Design, Ltd.) together with X-ray diffractometry (XRD).

Figure 2 shows the magnetic susceptibility of the deposit under a field of 0.002 T as a function of temperature. The closed and open circles represent the zero-field-cool susceptibility ($\chi_{ZFC}$) and the field-cool susceptibility ($\chi_{FC}$), respectively. $\chi_{ZFC}$ at 5 K is negative, and its amplitude is considerably large : $-1.6 \times 10^{-3}$ emu/g. With increasing the temperature, $\chi_{ZFC}$ increases gradually to a critical temperature ($T_c$) of 37 K, at which it becomes nearly equal to zero (see the inset). $\chi_{ZFC}$ remains in the vicinity of zero and does not show apparent temperature dependence from $T_c = 37$ K up to 65 K. Decreasing the temperature from 65 K under a field of 0.002 T, the field-cool magnetic susceptibility ($\chi_{FC}$) follows the $\chi_{ZFC}$ curve down to $T_c$. $\chi_{FC}$ drops to negative at $T_c$ and keeps on decreasing with lowering the temperature down to 5 K, at which $\chi_{FC}$ is about ten percent of $\chi_{ZFC}$. It is reasonable to interpret the diamagnetism observed in both $\chi_{ZFC}$ and $\chi_{FC}$ below $T_c$ as a consequence of the Meissner effect of a superconductor with $T_c = 37$ K in the deposit, because the considerably large amplitude of $\chi_{ZFC}$ at low temperatures is not explained by the inclusion of diamagnetic impurities.

Figure 3 shows the field dependence of magnetization measured on the same specimen of Fig.2 in the field ranging from -5 to 5 T at 4.5 K. With increasing the magnetic field from zero, the magnetization ($M_{4.5K}$) decreases steeply down to a bottom at 0.07 T (see the inset). $M_{4.5K}$ increases above 0.07 T with increasing the field, crosses the zero-axis, and continues to increase monotonously to 5 T, showing a tendency to saturation. With decreasing the field, $M_{4.5K}$ decreases to a bottom at 0.3 T. Below 0.3 T, $M_{4.5K}$ grows to a peak at the zero field and decreases again. The behavior of $M_{4.5K}$ in the third quadrant (both field and magnetization <0) is identical to that in the first quadrant (both field and magnetization >0). The monotonous increase of the magnetization above 0.3 T can be regarded as a contribution of some paramagnetic impurity because $M_{4.5K}$ at the fields above 0.3 T can be fitted well by a paramagnetic magnetization curve $M_{para}$ (H) : $M_{para}$ (H) = 0.237 B(J = 2.57 ; T = 4.5 K, H), where B(J; T, H) is the Brillouin function (the dotted curve in Fig.3). The hysteresis loop of the intrinsic magnetization defined by $M_{4.5K} - M_{para}$ turns out to be typical of a type-II superconductor. In particular, the minimum at 0.07 T in the virgin magnetization curve can be explained by penetration of magnetic flux into a type-II superconductor (see the inset).

Among all the reported compounds composed by combination of the constituent elements of the electrolyte (Mg, B, K, O, and Cl), $MgB_2$ is the only type-II superconductor with $T_c$ higher than 35 K. Thus, the results of the magnetic measurements suggest strongly that the deposit on the cathode contains $MgB_2$.

Figure 4 shows an XRD pattern (Cu K$\alpha$ radiation) on the cleaved face of the deposit. Four reflections are observed at $2\theta = 34.1°, 37.3°, 38.0°$ and 42.5$^o$ in the experimental profile (a) in addition to the 220 reflection of KCl at 2θ = 40.1$^o$. The reflections at 2θ = 34.1$^o$ and 42.5$^o$ are consistent with the 100 and 101 reflections of MgB$_2$ (reflections with the second and the first strongest intensity, respectively). The lattice constants calculated from the reflection positions are $a$ = 0.303(7) nm and $c$ = 0.361(2) nm, which agree with the reported lattice constants of MgB$_2$ ($a$ = 0.3086 nm and $c$ = 0.3524 nm)[1] within experimental errors. The rest of the reflections at 2θ = 37.3$^o$ and 38.0$^o$ could be assigned to no compound out of the constituent elements of the electrolyte, which may come from some impurity included in the starting materials or the constituents of the electrolysis cell.

The XRD pattern and the results of the magnetic measurements indicate that the deposit grown on the cathode contains the superconductive boride, MgB$_2$. From the amplitude of χ$_{ZFC}$ at 5 K, the fraction of MgB$_2$ in the deposit is estimated to be an order of a tens molar percent. The relatively low molar fraction of MgB$_2$ mainly results from a considerable amount of inclusion of the electrolyte constituent KCl in the deposit, as is seen in the XRD profile in Fig.4.

It is worth noting the formation condition of MgB$_2$ in the electrolysis system. Figure 5 shows the result of a voltanmetric measurement at the initial stage of electrolysis, at 600 $^o$C under applied voltages ranging from -3 to 3 V. The current-voltage profile in the positive and negative voltage regions represents the electrolysis on the anode and the cathode, respectively. The current increases almost monotonously with increasing the voltage in the positive voltage region, whereas an apparent kink is recognized in the negative voltage region at around -1.6 V. This suggests the existence of a threshold voltage $V_{th}$ of -1.6 V required for the formation of the deposit on the cathode. In fact, it was confirmed that no superconductive compound is obtained when the cathode voltage is higher than $V_c$.

In summary, electrolysis was performed on a fused mixture of MgB$_2$O$_4$, KCl, and MgCl$_2$. The results of the magnetic measurements and XRD indicate that the resultant deposit on the cathode contains MgB$_2$. This shows a synthesis route of MgB$_2$ with the following advantages : simple installation, low cost of the starting compounds, and a moderate reaction condition.

The authors would like to thank Motoharu Imai of NIMS for his useful comments on the paper.

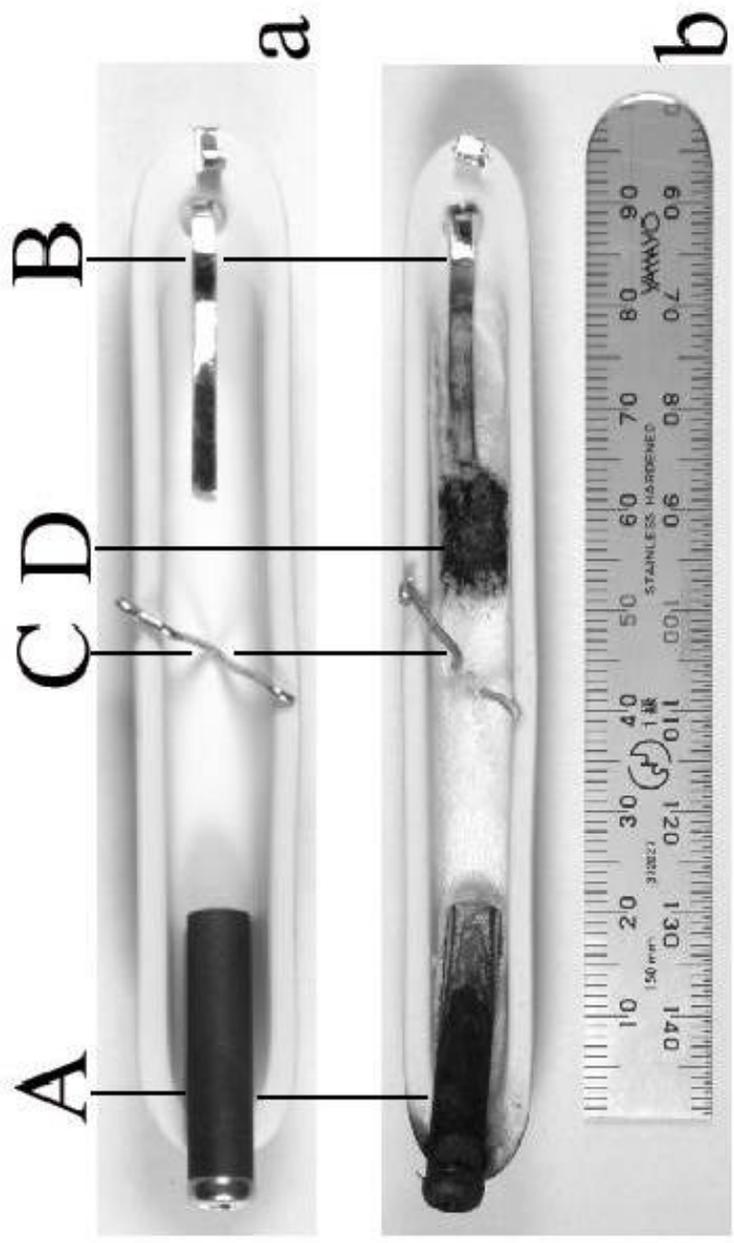

Figs.1.
Top views of the electrolysis cell before (a) and after (b) the electrolysis procedure.

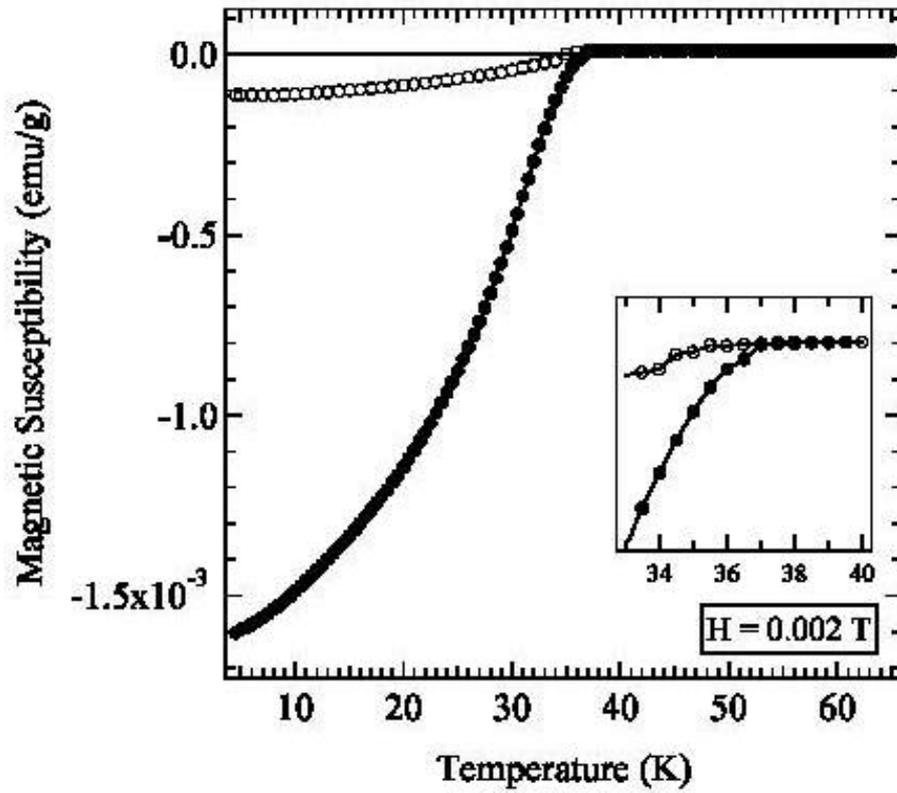

Fig.2.
Temperature dependence of magnetic susceptibility of the deposit grown on the cathode.    Closed and open circles represent $\chi_{ZFC}$ and $\chi_{FC}$, respectively. The inset shows a close-up view at around $T_c = 37$ K.

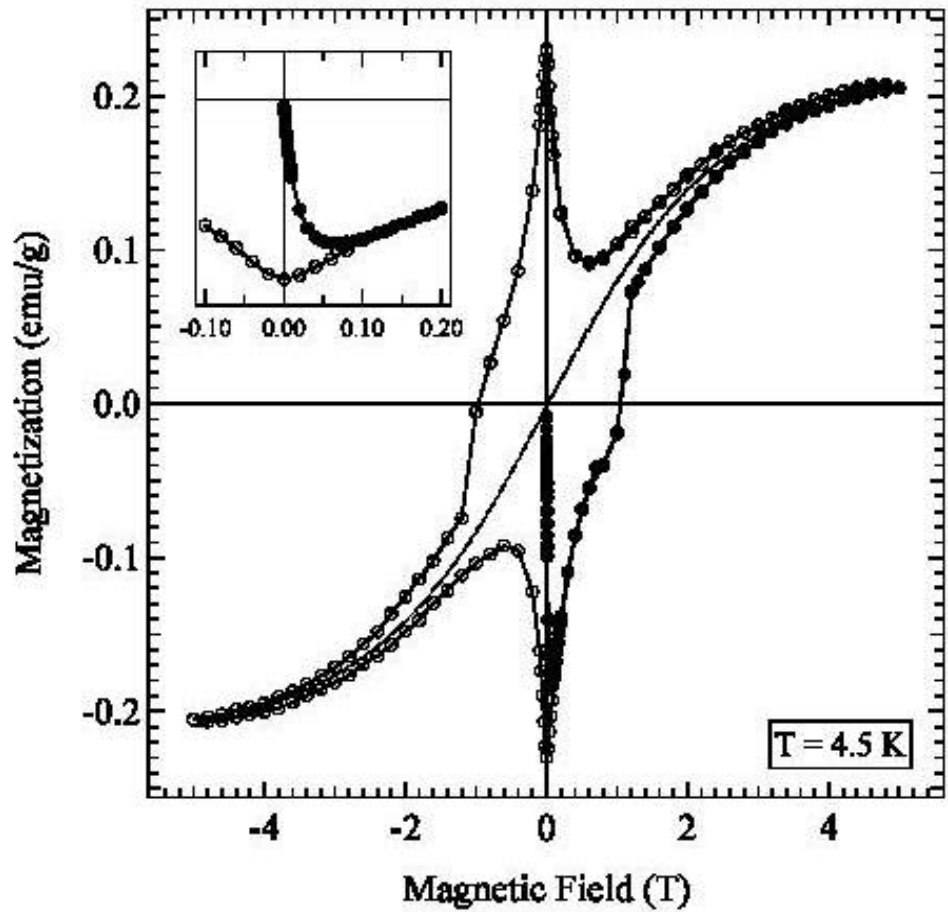

Fig.3.
Field dependence of the magnetization of the deposit at 4.5 K.    The virgin magnetization process is denoted as closed circles.    The inset shows a close-up view near around the zero-field.    The Brillouin function fitting to the experimental data ($M_{para}(H)$) is shown as a dotted curve (refer to the text).

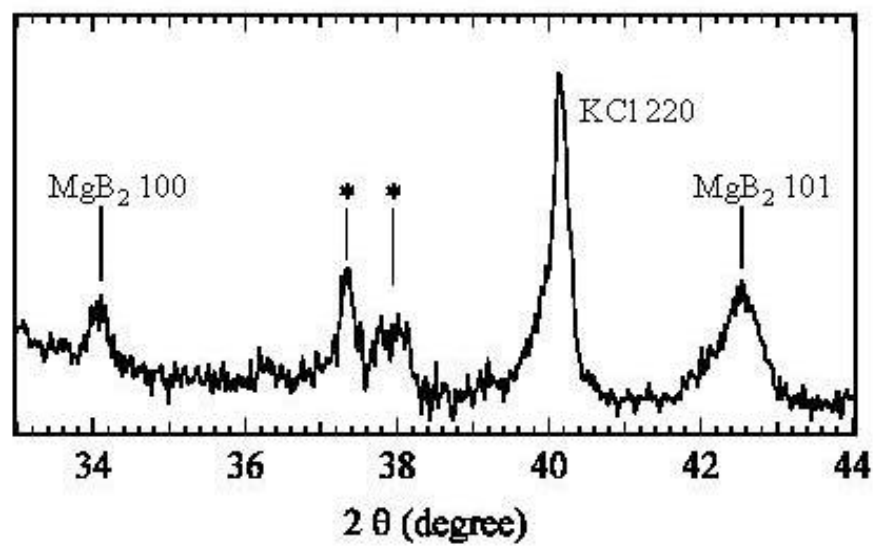

Fig.4.
XRD profile of the deposit grown on the cathode.    An assignment of the reflections is represented.

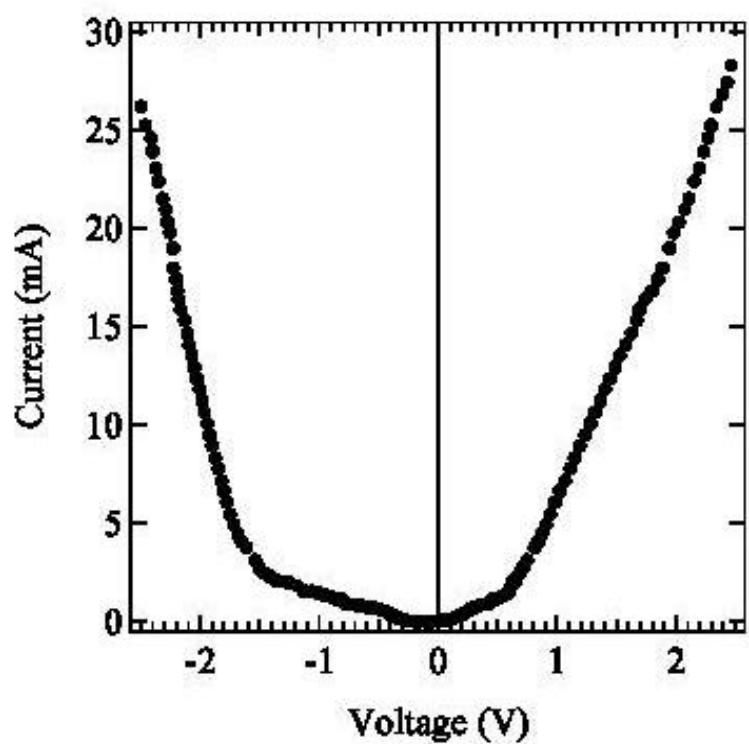

Fig.5.
Voltanmetry curve at the initial stage of electrolysis.